\documentclass[12pt,a4paper]{article}
\usepackage{fullpage}
\usepackage{graphicx}
\usepackage[latin1]{inputenc}
\usepackage[super,comma]{natbib}
\usepackage{psfrag}

\usepackage{setspace}

\title{Scientific citations in Wikipedia}
\author{Finn Årup Nielsen}

\begin{document}
\maketitle
\begin{center}
  {\footnotesize
    Lundbeck Foundation Center for Integrated Molecular Brain
    Imaging; \\
    Informatics and Mathematical Modelling,
    Technical University of Denmark, Lyngby, Denmark;  
    Neurobiology Research Unit, Copenhagen University Hospital 
    Rigshospitalet, Copenhagen, Denmark}
\end{center}


\begin{abstract}
  The Internet-based encyclopædia Wikipedia has grown to become one of 
  the most visited web-sites on the Internet.
  However, critics have questioned the
  quality of entries\cite{McHenryR2004FaithBased,DenningP2005Wikipedia},
  and an empirical study has shown Wikipedia to contain
  errors in a 2005 sample of science entries\cite{GilesJ2005Internet}.
  Biased coverage and lack of sources are among the ``Wikipedia
  risks''\cite{DenningP2005Wikipedia}.
  The present work describes a simple assessment of these
  aspects by examining the outbound links from Wikipedia articles to
  articles in scientific journals with a comparison against journal
  statistics from {\em Journal Citation Reports} such as impact factors.
  The results show an increasing use of structured citation markup and
  good agreement with the citation pattern seen in the scientific
  literature though with a slight tendency to cite articles in
  high-impact journals such as {\em Nature} and {\em Science}. 
  These results increase confidence in Wikipedia as an good information
  organizer for science in general.  
\end{abstract}

Wikipedia increases in popularity and will probably get further
importance for organization and dissemination of scientific research.
But how can the articles of this freely edited Internet-based
encyclopædia be trusted?

Inbound links can to some extent quantify the quality of a work, and
examples include Google's PageRank for web-pages and the impact factor
of scientific journals. 
The algorithms behind the PageRank and Kleinberg's
HITS\cite{KleinbergJon1999Authoritative} can be adapted to
Wikipedia\cite{BellomiF2005Network}, but it is not clear whether
high-scoring articles are also quality articles with respect to
content.
It has been suggested\cite{NeusA2001Managing,CrossT2006Puppy} 
that Wikipedia content surviving over a long period and many
edits may be deemed of high quality.
On the other hand studies have found that highly edited articles are
likely quality articles\cite{WilkinsonD2007Assessing}.
Other proposals for quality assessment use revision history to compute
a trust index for an article or an author reputation
index\cite{ZengH2006Computing,AdlerB2007ContentDriven}.
Another feature of an article that may correlate with article
quality is the amount of outbound citation to ``trusted'' material,
e.g., scientific articles. 
How prolific are these and does Wikipedia use them across scientific fields?
Critics have noted that Wikipedia may be biased on the corpus
level---leaned towards topics that interest the ``young and
Internet-savvy''---and a possible lack of sources has been
noted\cite{DenningP2005Wikipedia}.

Authors can include scientific references in Wikipedia by different
means, most simply, by listing them at the bottom of the article.
A more structured approach uses the {\tt <ref>} construct and the
{\em cite journal} template which allow for inline referencing and
consistent formatting. 
A user of the {\em cite journal} template needs to fill out the
appropriate bibliographic fields of the template, e.g., the fields for
the article title and the name of the journal.
The structured citation markup makes it relatively easy to extract
bibliographic information and ask: 
How well do the outgoing scientific citations in Wikipedia compare with
the citations seen between scientific journals?

To answer this question programs with regular expression matching
written in 
the Perl language 
extracted the journal titles from the {\em cite journal} templates in
all pages of the English Wikipedia obtained as the XML database dump
file.
A small list was setup to match the different variations of
journal titles, and then the total number of citations was counted for
each 
individual journal.
The {\em Journal Citation Reports} (JCR) for 2005 of {\em Thomson
  Scientific} provided statistics on citations between scientific
journals.

\begin{figure}
  \centering
  \psfrag{Kendall's tau}{Kendall's $\tau$}
  \psfrag{P-value}{$P$-value}

  \psfrag{Journals}{Journals}

  \psfrag{JCR total citations}{{\small JCR total citations}}
  \psfrag{JCR impact factor}{{\small JCR impact factor}}
  \psfrag{JCR articles}{{\small JCR articles}}
  \psfrag{JCR total citations x impact factor}{{\small JCR total
      citations $\times$ impact factor}}
  
  \includegraphics[width=0.95\textwidth]{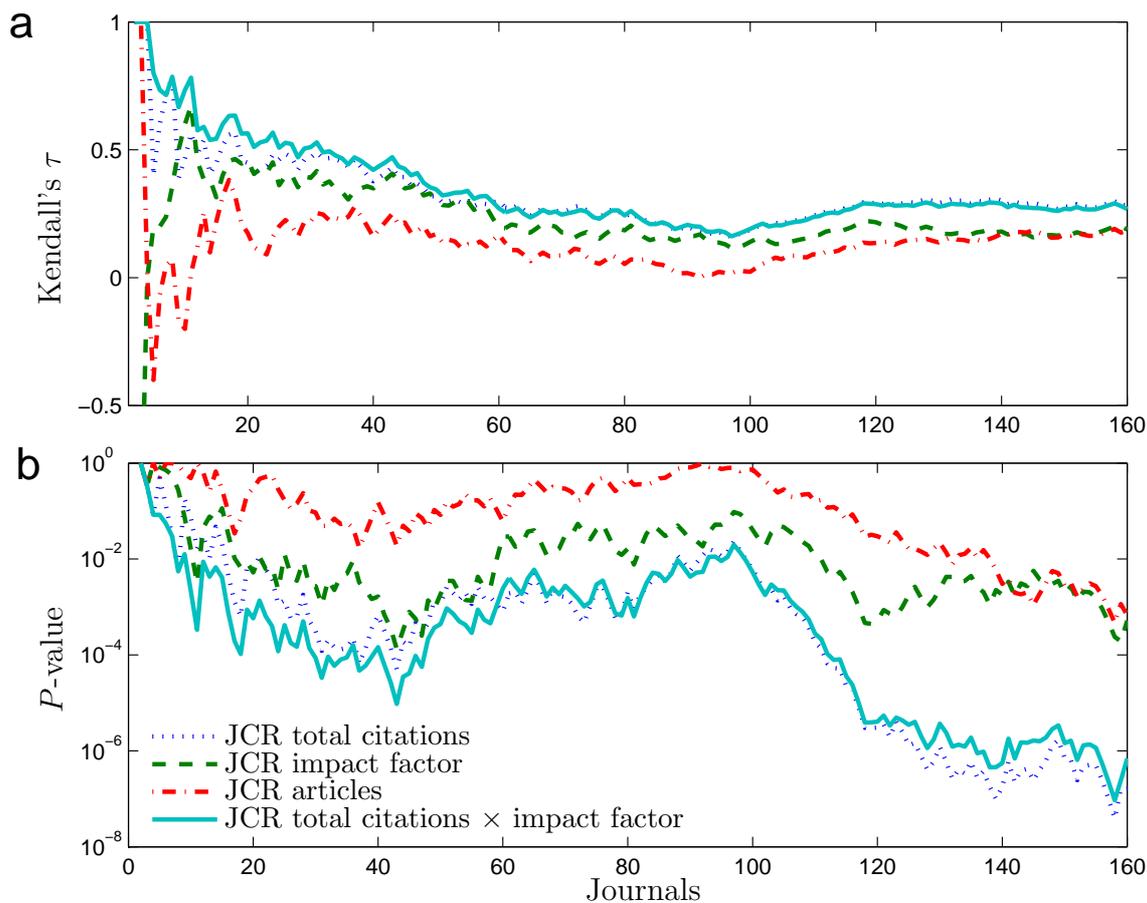}
  \caption{Correlations between citations to a journal from
    Wikipedia and from scientific journals.
    Kendall's rank correlation ({\bf a}) and its associated
    $P$-value ({\bf b}) as a function of the
    number of journals included in the test, e.g., 
    the value at 80 shows the correlation between Wikipedia citations
    and JCR numbers for the 80 most cited journals from Wikipedia.
    The number of citations from Wikipedia is compared with three
    series of numbers from JCR and one derived: The total
    citations to a journal, its impact 
    factors, the number of articles and the product of the total
    citations and impact factor.
  }
  \label{fig:correlationevolution}
\end{figure}

\begin{figure}[t]
  \psfrag{Wikipedia citations}{Wikipedia citations}
  \psfrag{ISI total citations x impact factor}{JCR total citations
    $\times$ impact factor}
  
  \centering
  \includegraphics[width=0.95\textwidth]{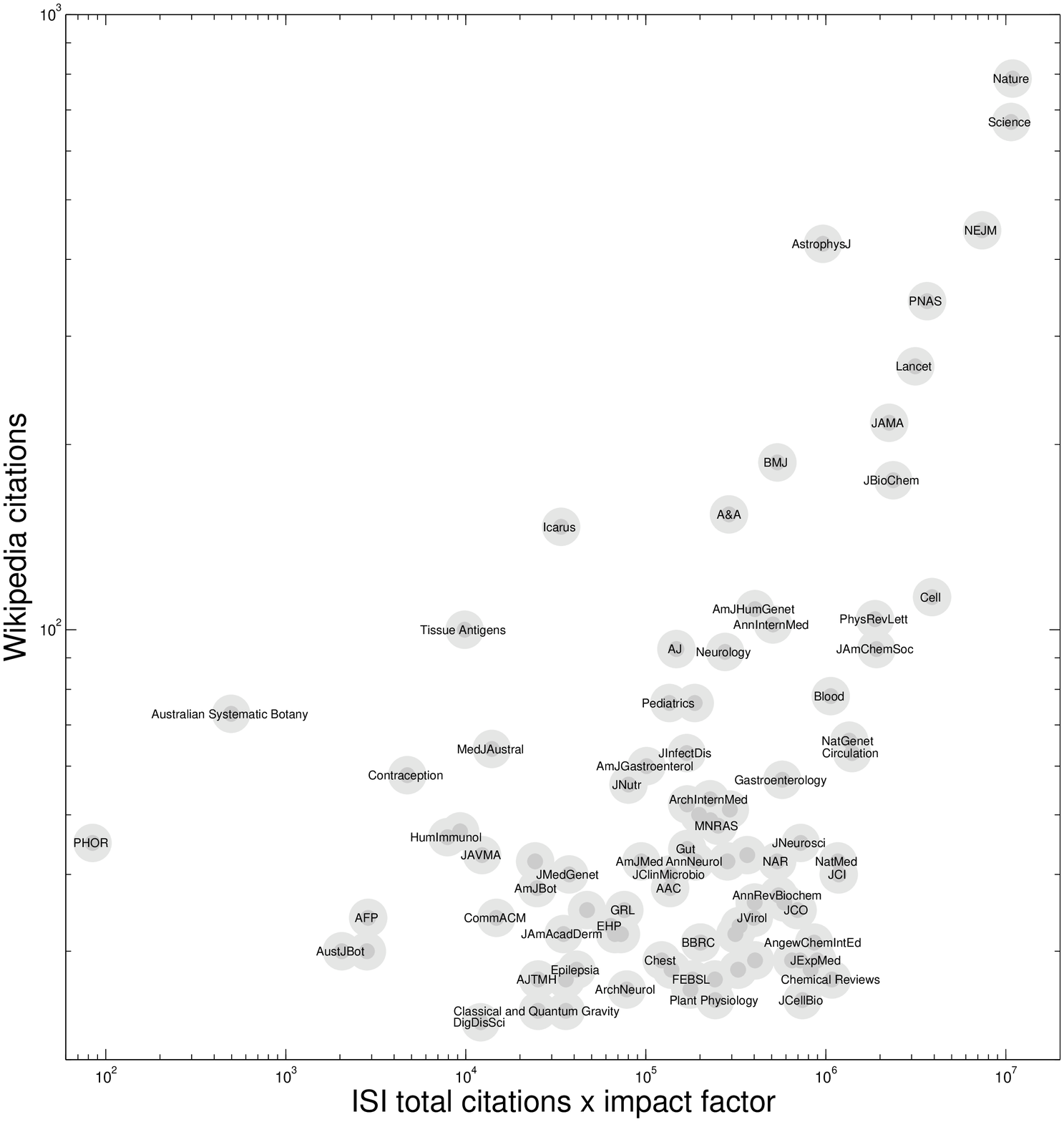}
  \caption{Comparison between citations from scientific
    journals and from Wikipedia.
    Scatter plot with each dot representing the target journal receiving
    the citations, and with one axis
    representing the number of citations from Wikipedia and the other the
    product of two numbers: JCR total citations and impact factor. 
    It indicates the 100 most Wikipedia referenced articles. The plot
    shows not all journal titles.
  } 
  \label{fig:scatter}
\end{figure}

The regular expression matched 30368 outbound citations from the 
{\em cite journal} template with the database dump for 
2 April 2007.
The summary statistics for the individual journals with the largest
number of inbound citations from Wikipedia showed {\em Nature} (787),
{\em Science} (669) and {\em New England 
  Journal of Medicine} (NEJM) (446) on the top (number of citations in
parenthesis).  
A number of astronomy journals received many citations: {\em
  The Astrophysical Journal} (424), {\em Astronomy \& Astrophysics}
(154), {\em Icarus, International Journal of Solar System Studies}
(147) and {\em The Astronomical Journal} (93).
Apart from NEJM other medical journals high on the list included {\em
  The Lancet} (268), {\em JAMA} (217), {\em British Medical Journal}
(187) and {\em Annals of Internal Medicine} (104).
Some newspapers and non-scientific journals also received
citations via the {\em cite journal} template with, e.g., 
{\em The New York Times} (69) among the most referenced. 
These non-scientific entries as well as journals such as {\em
  Scientific American} and {\em Physical
  Review} (that as a ``multivolume'' journal may be referenced in
several ways) were excluded and the rest of the values were
correlated against numbers obtained from JCR
(Fig.~1). 
The Wikipedia citation numbers showed high correlation with
the JCR's numbers for the total number of citations to a journal.
Wikipedia citation numbers correlated less with JCR impact factor and
the JCR's measure of numbers of articles in a journal. 
With 47.4 {\em Annual Review of Immunology} has the highest impact,
but because it publishes few articles it receives
relatively few citations both from scientific journals and from
Wikipedia (18).
The correlations depended on the number of journals included in the
test, with the largest correlation observed for the highly cited journals. 
It may simply reflect that journals with a small number of
citations make noisy and poor statistics.
In most cases the highest correlation could be obtained by
multiplying the total number of citation with the impact factor, 
i.e., Wikipedia authors slightly overcite high-impact journals
compared to JCR numbers.  
The high correlation among top-cited journals with this combined
number means that the 10 journals with the highest value of this
measure feature
among the 19 most Wikipedia-referenced journals.

When individual journals are examined 
Wikipedia citations to astronomy journals stand out
compared to the overall trend (Fig.~2). 
Also Australian botany journals received a considerable number of
citations, e.g., 
{\em Nuytsia} (101), in part due to concerted effort for the genus
{\em Banksia}, where several Wikipedia articles for {\em Banksia}
species have reached ``featured article'' status.
Computer and Internet-related journals do not get as many as one
would expect if Wikipedia showed bias towards fields for
the ``Internet-savvy''. 
{\em Communications of the ACM} (34) became the most referenced. 
Of the medical journals BMJ received relatively many Wikipedia
citations.
Authors cite more often freely available
articles\cite{LawrenceS2001Free}, 
and this may be particularly true for authors of the free encyclopædia.
Since BMJ's research articles are free the journal may gain extra
citations from this effect.

Citing Wikipedia as an authoritative source may be questionable with
the present state of review on Wiki\-pedia, and some universities have
even banned citations to Wiki\-pedia\cite{CohenN2007History}.
But when citations to trusted material support statements Wikipedia
may be valuable for background reading. 
The present number of structured outbound citations from Wikipedia dwarfs in
relation to the total number of scientific citations in the entire
scientific literature.
With this low number dedicated enthusiasts can influence the
statistics making relatively few edits, cf.\ Australian botany.
However, the use of the {\em cite journal} template has grown from
zero in 
February 2005 when first introduced, to 19066 in November
2006, 
24656 in February 2007, to a total of 30368 citations in April 2007. 
Reference management software (Zotero) now includes functionality for
handling Wikipedia citations. 
Thus use of structured scientific citations in Wikipedia will very
likely continue to grow and increasingly benefit researchers that look
for well-organized pointers to original research.


\begin{thebibliography}{10}

\bibitem{McHenryR2004FaithBased}
McHenry, R.
\newblock The faith-based encyclopedia.
\newblock {\em TCS Daily}{ \bf } November  (2004).
\newblock http://www.techcentralstation.com\-/111504A.html.
\newblock Accessed 2006-03-05.

\bibitem{DenningP2005Wikipedia}
Denning, P., Horning, J., Parnas, D., and Weinstein, L.
\newblock Wikipedia risks.
\newblock {\em Comm. ACM}{ \bf 48}, 152 December  (2005).

\bibitem{GilesJ2005Internet}
Giles, J.
\newblock Internet encyclopaedias go head to head.
\newblock {\em Nature}{ \bf 438}, 900--901 December  (2005).

\bibitem{KleinbergJon1999Authoritative}
Kleinberg, J.~M.
\newblock Authoritative sources in a hyperlinked environment.
\newblock {\em J. ACM}{ \bf 46}, 604--632 (1999).

\bibitem{BellomiF2005Network}
Bellomi, F. and Bonato, R.
\newblock Network analysis of {Wikipedia}.
\newblock In {\em Proceedings of {Wikimania} 2005 --- The First International
  {Wikimedia} Conference},  (2005).

\bibitem{NeusA2001Managing}
Neus, A.
\newblock Managing information quality in virtual communities of practice.
\newblock In {\em Proceedings of the 6th International Conference on
  Information Quality at {MIT}, }Pierce, E. and Katz-Haas, R., editors (Sloan
  School of Management, Boston, MA, 2001).

\bibitem{CrossT2006Puppy}
Cross, T.
\newblock Puppy smoothies: Improving the reliability of open, collaborative
  wikis.
\newblock {\em First Monday}{ \bf 11}(9) September  (2006).


\bibitem{WilkinsonD2007Assessing}
Wilkinson, D.~M. and Huberman, B.~A.
\newblock Assessing the value of cooperation in {{\em Wikipedia}}.
\newblock {\em First Monday}{ \bf 12}(4) April  (2007).


\bibitem{ZengH2006Computing}
Zeng, H., Alhossaini, M., Ding, L., Fikes, R., and McGuinness, D.~L.
\newblock Computing trust from revision history.
\newblock In {\em Proceedings of the 2006 International Conference on Privacy,
  Security and Trust},  October  (2006).

\bibitem{AdlerB2007ContentDriven}
Adler, B.~T. and de~Alfaro, L.
\newblock A content-driven reputation system for the {Wikipedia}.
\newblock In {\em Proceedings of the Sixteenth International World Wide Web
  Conference ({WWW2007}) {May} 8-12, 2007 {Banff}, {Alberta}, {CANADA}},  261+,
   (2007).

\bibitem{LawrenceS2001Free}
Lawrence, S.
\newblock Free online availability substantially increases a paper's impact.
\newblock {\em Nature}{ \bf 411}, 521  (2001).

\bibitem{CohenN2007History}
Cohen, N.
\newblock A history department bans citing {Wikipedia} as a research source.
\newblock {\em New York Times}{ \bf } February  (2007).

\end{thebibliography}

\vspace{5mm}

\noindent {\bf Acknowledgements.} 
I thank D. Balslev, R. Jesus and L.K. Hansen for discussions and the
Lundbeck Foundation for support.

\end{document}